\documentclass[superscriptaddress,twocolumn,aps,prl,showpacs,amsmath,amssymb]{revtex4-1}

\usepackage[usenames]{color}
\usepackage{graphicx}
\usepackage{amsmath,stackrel}

\begin{document}

\title{Efimov Physics in Quenched Unitary Bose Gases}

\author{Jos\'e P. D'Incao}
\affiliation{JILA, University of Colorado and NIST, Boulder, Colorado 80309-0440, USA}
\affiliation{Department of Physics, University of Colorado, Boulder, Colorado 80309-0440, USA}
\author{Jia Wang}
\affiliation{Centre for Quantum and Optical Science, Swinburne University of Technology, Melbourne 3122, Australia}
\author{V. E. Colussi}
\affiliation{JILA, University of Colorado and NIST, Boulder, Colorado 80309-0440, USA}
\affiliation{Department of Physics, University of Colorado, Boulder, Colorado 80309-0440, USA}
\affiliation{Eindhoven University of Technology, PO Box 513, 5600 MB Eindhoven, The Netherlands}

\begin{abstract}
We study the impact of three-body physics in quenched unitary Bose gases, focusing on the role of the Efimov effect. 
Using a local density model, we solve the three-body problem and determine three-body decay rates 
at unitary, finding density-dependent, log-periodic Efimov oscillations, violating the expected continuous scale-invariance in the system.  
We find that the breakdown of continuous scale-invariance, due to Efimov physics, manifests also in the earliest stages of evolution 
after the interaction quench to unitarity, where we find the growth of a substantial population of Efimov states for densities in which the 
interparticle distance is comparable to the size of an 
Efimov state. This agrees with the early-time dynamical growth of three-body correlations at unitarity 
[Colussi {\em et al}., Phys. Rev. Lett. 120, 100401 (2018)]. By varying the sweep rate away from 
unitarity, we also find a departure from the usual Landau-Zener analysis for state transfer when the system is allowed to evolve
at unitarity and develop correlations. 
\end{abstract}

\maketitle 

Over the past few years, the study of strongly interacting Bose gases 
has greatly intensified due to the experimental advances with ultracold atoms 
\cite{fletcher2013PRL,rem2013PRL,makotyn2014NTP,eismann2016PRX,chevy2016JPB,fletcher2017Sci,laurent2017PRL,
klauss2017PRL,eigen2017PRL,fletcher2018ARX},
unraveling universal properties and other intriguing phenomena 
\cite{ho2004PRL,diederix2011PRA,weiran2012PRL,jiang2014PRA,laurent2014PRL,yin2013PRA,sykes2014PRA,
rossi2014PRA,smith2014PRL,braaten2011PRL,werner2012PRA,piatecki2014NTC,kira2015NTC,
barth2015PRA,corson2015PRA,tommaso2016PRL,yin2016PRA,jiang2016PRA,ding2017PRA,colussi2018PRL,sze2018PRA,blume2018PRA}.
Although ultracold quantum gases have extremely low densities, $n$, the unique ability to control 
the strength of the interatomic interactions ---characterized by the $s$-wave scattering length, $a$--- via Feshbach resonances 
\cite{chin2010RMP} allows one to probe the {\em unitary regime} ($n|a|^3\gg1$), where the probability for collisions 
can reach unity value and the system becomes non-perturbative.
In contrast to their fermionic counterparts \cite{giorgini2008RMP,holland2001PRL,kokkelmans2002PRA}, unitary Bose gases are susceptible to 
fast atomic losses \cite{dincao2004PRL} that can prevent the system from reaching
equilibrium. 
In Ref. \cite{makotyn2014NTP}, a quench of the interactions from weak
to strong allowed for the study of the dynamical evolution and equilibration of the unitary Bose gas, thanks to the surprisingly slow 
three-body decay rates \cite{sykes2014PRA}.
By making the unitary regime accessible, this new quenched scenario opened up intriguing
ways to study quantum few- and many-body non-equilibrium dynamics in a controlled manner
\cite{polkovnikov2011RMP,rigol2008NT,dziarmaga2010AP,gring2012Sci,cazalilla2012PRE,zurec2005PRL}. 

Our understanding of how correlations evolve and subsequently equilibrate in quenched unitary Bose gases is evolving
as recent experiments probe physics in this regime 
\cite{fletcher2013PRL,rem2013PRL,makotyn2014NTP,eismann2016PRX,chevy2016JPB,fletcher2017Sci,klauss2017PRL,eigen2017PRL,fletcher2018ARX}. 
Most of the current theoretical approaches, however, are based on the two-body physics alone, leaving  
aside the three-body Efimov 
physics \cite{efimov1979SJNP,efimov1973NPA,braaten2006PRep,wang2013Adv,naidon2017RPP,dincao2018JPB,greene2017RMP}.
In particular, at unitarity ($|a|=\infty$), although no weakly bound two-body state exists, 
an infinity of Efimov states form.
Critical aspects such as the three-body loss rates and dynamical formation of Efimov state populations remain 
unexplored within the non-equilibrium scenario of quenched unitary Bose gases.

In this Letter, we explore various aspects related to the three-body physics in quenched unitary Bose gases.  
We solve the three-body problem using a simple local model, incorporating density effects through a local harmonic trap and 
describing qualitatively Efimov physics embedded in a larger many-body system.
Within this model, we determine loss rates at unitarity that display 
density-dependent, log-periodic oscillations due to Efimov physics. We also analyze the dynamical formation  
of Efimov states when the quenched system is held at unitarity and then swept away to weak interactions. 
This scheme was recently implemented for an ultracold gas of $^{85}$Rb atoms \cite{klauss2017PRL},
where a population of Efimov states in a gas phase was observed for the first time. 
Our present study analyzes such dynamical  effects and demonstrates that for densities where the 
interparticle distance is comparable to the size of an Efimov state, their population is enhanced. 
This is consistent with a recent theoretical study on the early-time dynamical growth of three-body correlations 
\cite{colussi2018PRL}, providing additional evidence for the early-time violation of the universality hypothesis for the 
quenched unitary Bose gases \cite{ho2004PRL}. By studying the dependence of the populations on the sweep time, 
we find a departure from the usual Landau-Zener model of the state formation as the system evolves at unitarity 
and develops correlations.

Within the adiabatic hyperspherical representation \cite{suno2002PRA,dincao2005PRA,wang2011PRA,wang2012PRL}, 
the total three-body wave function for a given state $\beta$ is decomposed as 
\begin{align}
\psi_\beta(R,\Omega)=\sum_{\nu}F_{\beta\nu}(R)\Phi_{\nu}(R;\Omega),
\end{align}
where $\Omega$ collectively represents the set of all hyperangles, describing the internal motion and overall rotations of the system, 
and the hyperradius, $R$, gives the overall system size. The channel functions $\Phi_\nu$ are eigenstates of the 
hyperangular part of the Hamiltonian (including all interatomic interactions) whose eigenvalues are the hyperspherical
potentials, obtained for fixed values of $R$. Bound and scattering properties of the system are determined by solving the
three-body hyperradial Schr\"odinger equation,
\begin{align}
&\left[-\frac{\hbar^2}{2\mu}\frac{d^2}{dR^2}+W_{\nu}(R)\right]F_{\nu}(R)\nonumber\\
&~~~~~~+\sum_{\nu\ne\nu'}W_{\nu\nu'}(R)F_{\nu'}(R)=EF_{\nu}(R),\label{Schro}
\end{align}
where $\mu=m/\sqrt{3}$ is the three-body reduced mass and $\nu$ an index that includes all 
quantum numbers necessary to characterize each channel. The hyperradial motion is then described
by Eq.~(\ref{Schro}) and is governed by the effective three-body potentials $W_\nu$ and nonadiabatic 
couplings $W_{\nu\nu'}$. 
In our model, we assume atoms interact via a Lennard-Jones potential, 
$v(r)=-C_6/r^6(1-\lambda^6/r^6)$, where $C_6$ is the dispersion coefficient \cite{chin2010RMP}, and 
$\lambda$ is a parameter adjusted to give the desired value of $a$, tuned such 
that only a single $s$-wave dimer can exist \cite{SupMat}. 
The correct three-body parameter \cite{wang2012PRL,naidon2014PRA,naidon2014PRL}, found in terms of the van der Waals length 
$r_{\rm vdW}=(mC_6/\hbar^2)^{1/4}/2$ \cite{chin2010RMP}, is naturally built into this potential model, 
providing a more realistic description of the problem.

\begin{figure}[htbp]
\includegraphics[width=3.in,angle=0,clip=true]{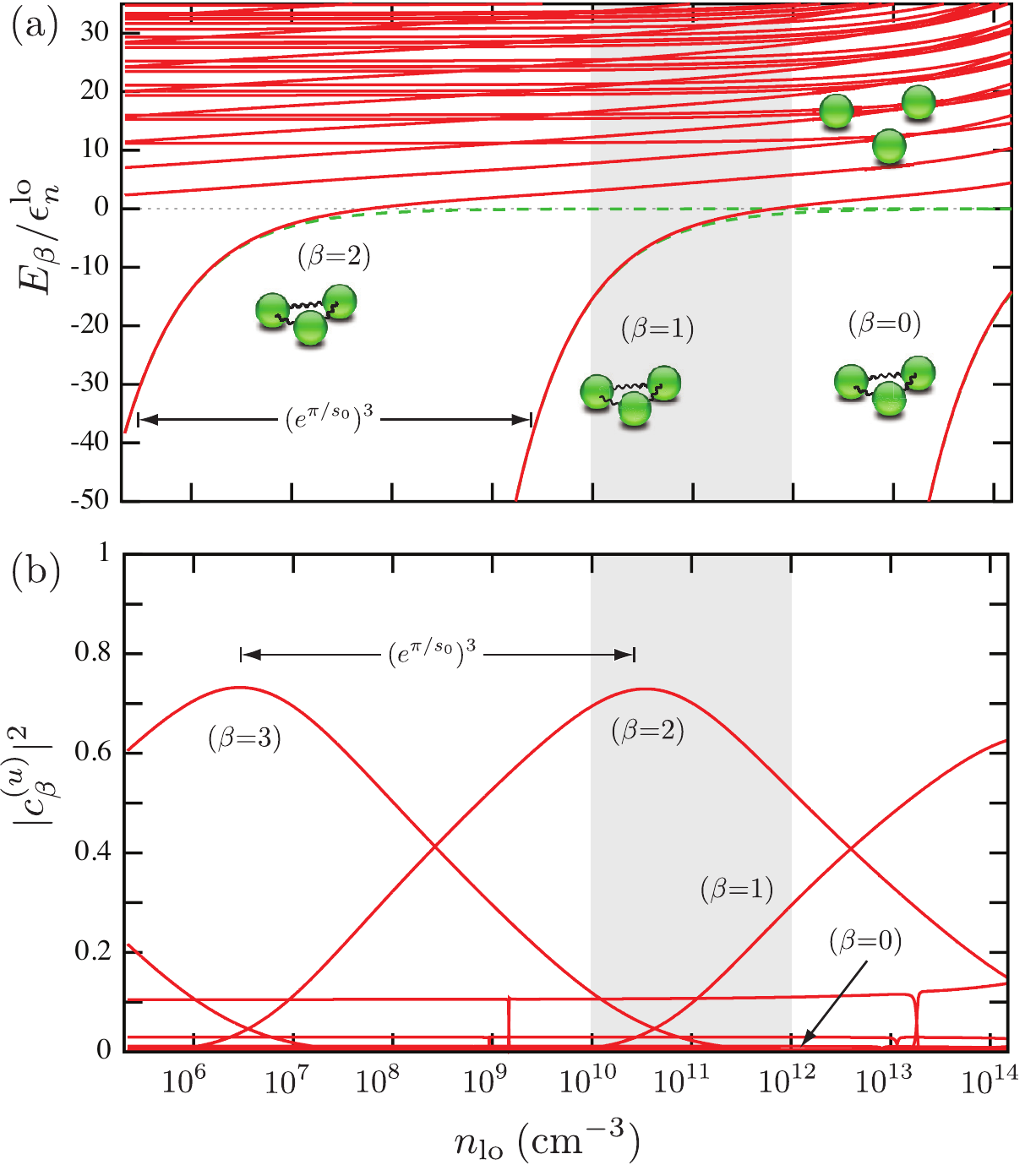}
\caption{(a) Energies, $E_{\beta}$ [in units of $\epsilon_{n}^{\rm lo}=\hbar^2(6\pi^2{n_{\rm lo}})^{2/3}/2m$], 
of three-identical bosons at $|a|=\infty$ as a function of the local density, $n_{\rm lo}$.
The dashed green lines represent the free space Efimov state energies. 
(b) The corresponding population of three-body states, $|c_{\beta}|^2$, for quenching to unitarity. 
The shaded region marks the range of densities where we have studied the dynamical formation of
Efimov states. 
}
\label{EnTnCn}
\end{figure}

In order to qualitatively incorporate density effects in our calculations, we introduce a {\em local} 
harmonic confinement whose properties are determined from the average atomic density, $\langle n\rangle$ 
\cite{goral2004JPB,borca2003NJP,stecher2007PRL,sykes2014PRA}. 
This allows us to connect local few-body properties to density-derived scales of the gas, including the 
system's energy, $\epsilon_{n}=\hbar^2(6\pi^2\langle{n}\rangle)^{2/3}/2m$, 
length scales, $k_n^{-1}=\hbar/(2m\epsilon_n)^{1/2}$, and time scales, 
$t_n=\hbar/\epsilon_n$. 
In the hyperspherical representation, local harmonic confinement is achieved by adding a 
hyperradial harmonic potential \cite{werner2006PRL,blume2012RPP},
\begin{align}
V_{\rm ho}(R)=\frac{\mu\:\omega_{\rm ho}^2}{2}R^2,\label{vho}
\end{align}
to the effective potentials $W_{\nu}$ in Eq.~(\ref{Schro}). Here, 
$\omega_{\rm ho}=\hbar/ma_{\rm ho}^2$ is the trapping frequency and $a_{\rm ho}$ is the
oscillator length. 
A priori, there is no unique way to relate the harmonic confinement in our model to the atomic density. 
Nevertheless, as shown in Refs.~\cite{goral2004JPB,borca2003NJP,stecher2007PRL,sykes2014PRA,colussi2018PRL,corson2015PRA},
calibrating the local trapping potential [Eq.~(\ref{vho})] to match the few-body density with the interparticle spacing 
($\sim \langle n\rangle^{-1/3}$) qualitatively describes the larger many-body system for time scales shorter than 
$t_{\rm ho}=1/\omega_{\rm ho}$.
Here, we relate the local atomic density, $n_{\rm lo}$, and local trapping potential by 
\begin{align}
n_{\rm lo}&=\bigg[\frac{4\pi}{3}\big\langle\psi_i\big|R^3\big|\psi_i\big\rangle\bigg]^{-1}\propto\frac{1}{a_{\rm ho}^3},
\label{nlink}
\end{align}
where $\psi_i$ is the three-body wave function of the lowest trap state in the regime of weak interactions.
The results of this Letter were obtained using $\psi_i$ relevant for the $^{85}$Rb experiment \cite{klauss2017PRL}, 
in which the pre-quench, initial state corresponds to $a\approx150a_0$.

In free space, and in the absence of a background gas, the energies of Efimov states at unitarity, $E_{\rm 3b}$,
accumulate near the free-atom threshold ($E=0$), and their corresponding sizes, $R_{\rm 3b}$, 
increase according to the characteristic log-periodic geometric scaling \cite{braaten2006PRep}:
\begin{align}
E_{\rm 3b}^{(j)}=-\frac{\hbar^2\kappa_*^2/m}{(e^{\pi/s_0})^{2j}}\mbox{~~and~~}
R_{\rm 3b}^{(j)}=\frac{(1+s_0^2)^{\frac{1}{2}}}{({3}/{2})^{\frac{1}{2}}\kappa_*}(e^{\pi/s_0})^j,\label{E3bR3b}
\end{align}
where $j$=0, 1, ..., labels each Efimov state according to their excitation, $\kappa_*\approx0.226/r_{\rm vdW}$ is the 
three-body parameter ~\cite{wang2012PRL}, and $e^{\pi/s_0}\approx22.7$ is Efimov's geometric factor 
for identical bosons.
In the unitary Bose gas, however, one expects that only Efimov states with binding energies larger than $\epsilon_n$, and sizes 
smaller than $k_n^{-1}$ are insensitive to the background gas and can exist in their free-space form. 
Otherwise, Efimov states should be sensitive to the background gas, represented here by a local trapping potential.
To illustrate this sensitivity within our model, the energy levels, $E_{\beta}$, of three-identical bosons at $|a|=\infty$ as a function 
of $n_{\rm lo}$ are shown in Fig.~\ref{EnTnCn}(a), displaying geometric scaling as $n_{\rm lo}$ is increased by a factor 
$(e^{\pi/s_0})^3\approx1.17\times10^4$. Within our model, as the energy of a Efimov state approaches $\epsilon_{n}^{\rm lo}$ its value is shifted 
away from its value in free-space [green dashed lines in Fig.~\ref{EnTnCn}(a)].
In order to describe loss processes within our model, we have also provided a finite width (lifetime) for 
the states, $\Gamma_{\beta}$ ($\tau_\beta=\hbar/\Gamma_\beta$), 
adjusted to reproduce the known behavior of the Efimov physics in $^{85}$Rb \cite{wild2012PRL} 
---see Supplementary Material \cite{SupMat}.

\begin{figure}[htbp]
\includegraphics[width=3.2in,angle=0,clip=true]{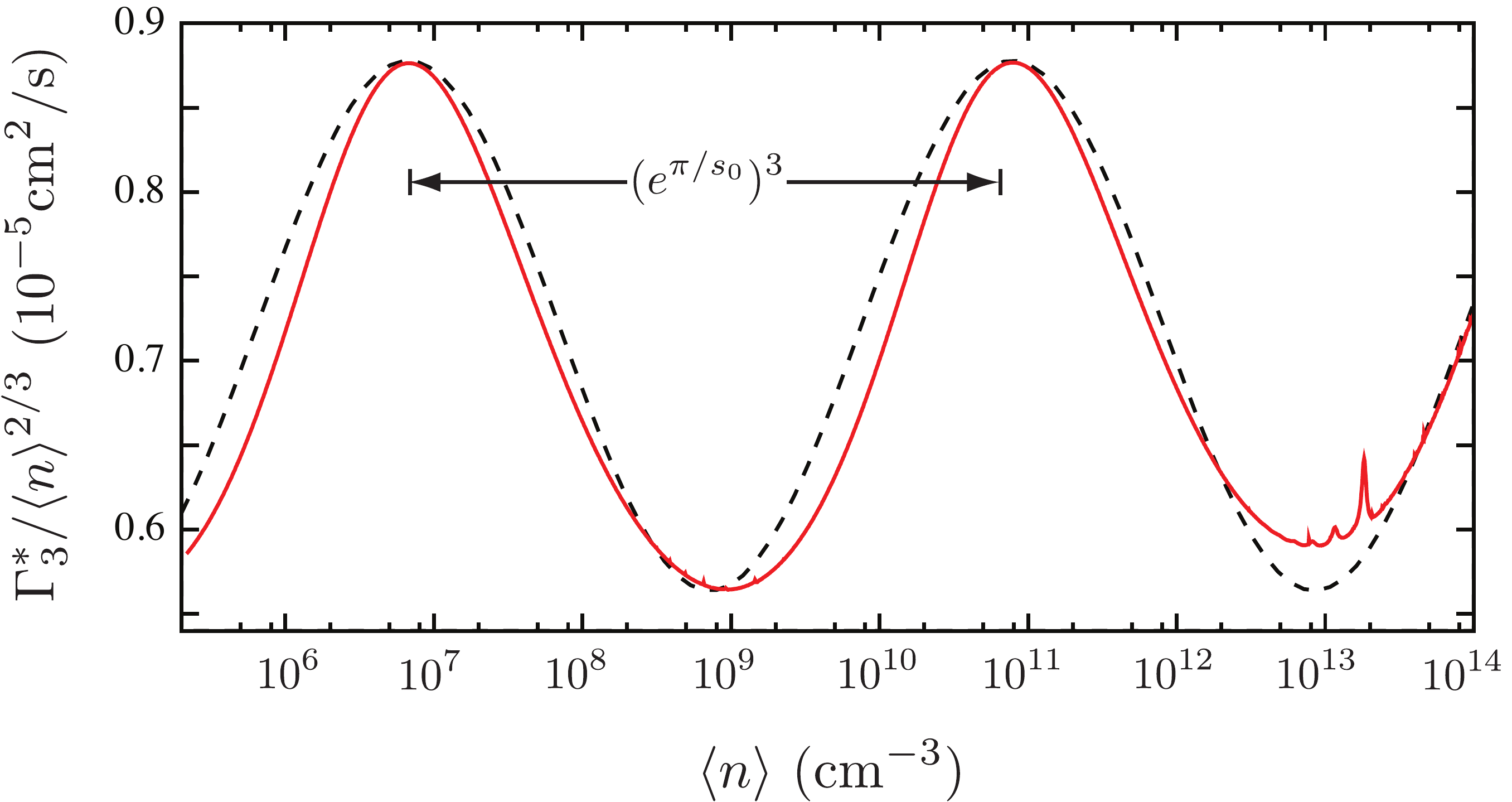}
\caption{Three-body decay rate, $\Gamma^*_3$, for $^{85}$Rb at $|a|=\infty$ as a function of the average 
density, $\langle n\rangle$. This figure display the $\langle n^{2/3}\rangle$ scaling of the decay rate as well as the appearance of log-periodic 
oscillations associated with Efimov physics. 
Resonances at higher $\langle n\rangle$ occur when the initial state is degenerate with one of the possibly weakly-coupled final states in our model.
Dashed black curve is the fitting function in Eq.~(\ref{GammaA}).}
\label{Gamma}
\end{figure}

Besides the three-body eigenenergies, our model also provides wave functions that determine various properties
of the quenched system. After quenching to the unitary regime, the time evolution of the three-body system is described 
by the projected three-body wave function,
\begin{align}
\Psi(R,\Omega,t)=\sum_\beta c_{\beta}^{(u)}\psi_\beta^{(u)}(R,\Omega)e^{-i(E_{\beta}-i\Gamma_{\beta}/2)t/\hbar},\label{evolve}
\end{align}
which is a superposition of states at unitary, $\psi_{\beta}^{(u)}$,
with coefficients determined from their overlap with the initial state, 
$c_{\beta}^{(u)}=\langle\psi^{(u)}_\beta|\psi_i\rangle$.
Within our local model, however, the wave function in Eq.~(\ref{evolve}) can only be expected to qualitatively represent the actual 
many-body system for $t<t_n$, since beyond this time scale genuine many-body effects should become important.
Figure \ref{EnTnCn}(c) shows the population at unitarity, $|c_{\beta}|^2$, for various states as a function of $n_{\rm lo}$.
We observe that the population of a given state becomes substantial when its energy or size 
is in the vicinity of the density-derived scales of the unitary Bose gas ($\epsilon_n$ and $k^{-1}_n$, respectively). 

\begin{figure*}[htbp]
\includegraphics[width=6.8in,angle=0,clip=true]{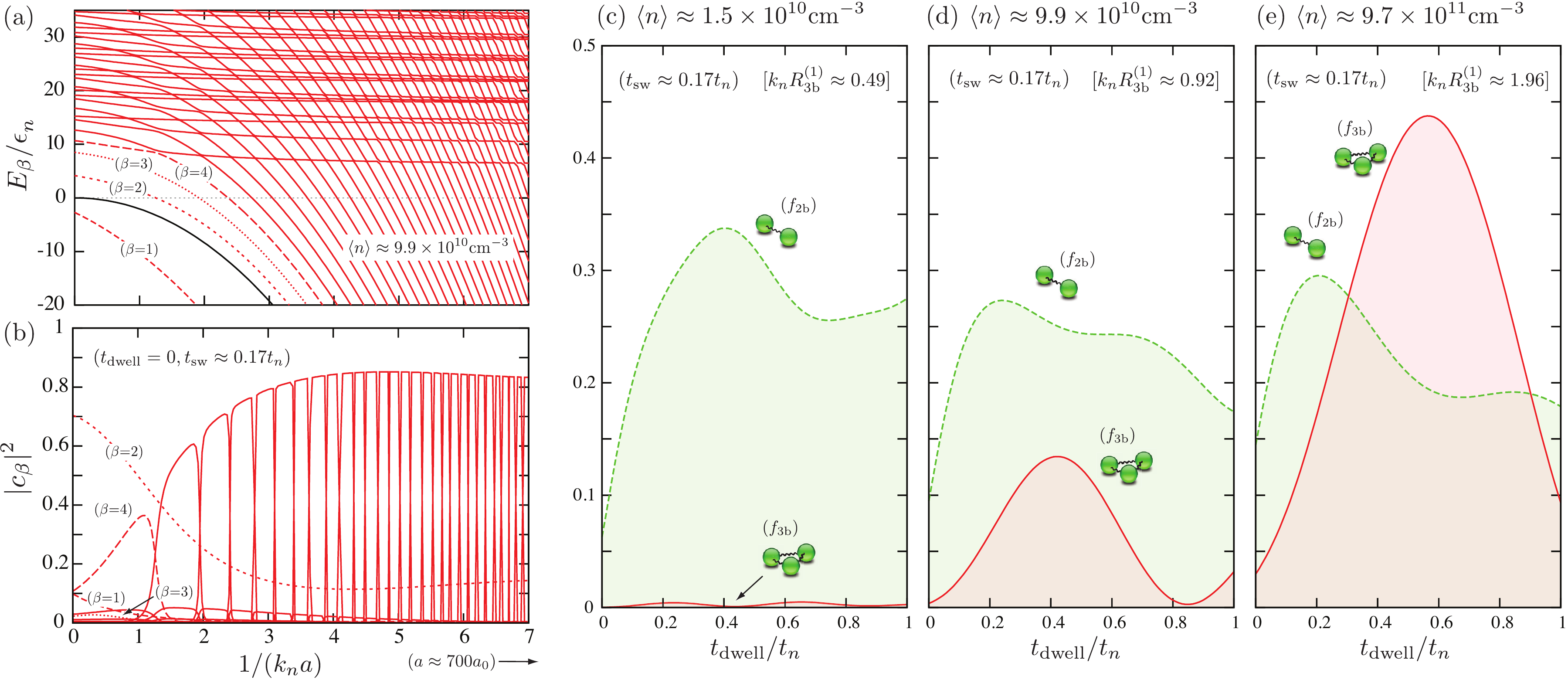}
\caption{(a) Three-body energy spectrum for $\langle n\rangle\approx9.9\times10^{10}$cm$^{-3}$ within a range of $a$ relevant for the 
$^{85}$Rb experiment. Black solid line corresponds to the energy of the (free-space) diatomic state 
($\hbar^2/ma^2$). (b) Corresponding change of population during the field sweep ($t_{\rm sw}\approx0.17t_n$) obtained immediately 
after the quench ($t_{\rm dwell}=0$). (c)-(e) Population fraction of diatomic and Efimov states formed as a function of $t_{\rm dwell}$ illustrating the
enhancement of Efimov state formation as the density approaches the characteristic value 
$\langle n_1^*\rangle\approx5.2\times10^{10}$cm$^{-3}$ [Eq.~(\ref{nc})] or, equivalently, when $k_nR_{\rm 3b}\approx0.78(2)$ \cite{colussi2018PRL}. 
}
\label{RampOut}
\end{figure*}

The above results suggest that Efimov physics may manifest in the density 
dependence of relevant observables since the atomic density sets the energy and length scales of the gas.
We first focus on the three-body decay rates, which can be simply evaluated within our model from \cite{sykes2014PRA}
\begin{align}
\Gamma_{3}^*=-\lim_{t\rightarrow0}\frac{\dot{n}(t)}{n(t)}
=\sum_\beta |c_{\beta}^{(u)}(n_{\rm lo})|^2\frac{\Gamma_{\beta}(n_{\rm lo})}{\hbar},\label{GammaF}
\end{align}
where $n(t)=n_{\rm lo}|\langle\Psi(R,\Omega,t)|\Psi(R,\Omega,t)\rangle|$. 
In a local density model, the decay rate in Eq.~(\ref{GammaF}) is averaged over the 
local density $n_{\rm lo}$. 
Using a Thomas-Fermi density profile, our numerical calculations
indicate that the averaged rate is well approximated (within no more than 4\%) by 
replacing $n_{\rm lo}$ with the corresponding average density $\langle n\rangle$ 
in Eq.~(\ref{GammaF}).
Our results for the three-body decay rate for the quenched unitary Bose gas are shown
in Fig. \ref{Gamma}, covering a broad range of densities. 
We find the expected $\langle n^{2/3}\rangle$ scaling \cite{werner2012PRA,smith2014PRL,klauss2017PRL} but 
also log-periodic oscillations that originate from the increase of an Efimov state population whenever its binding energy 
is comparable to $\epsilon_n$ (see Fig.~\ref{EnTnCn}).
We fit these oscillations as
\begin{eqnarray}
\Gamma_3^*\approx\eta\frac{\hbar}{m}\left[A+B\sin^2\Big(s_0\ln\scalebox{1.1}{$\frac{\langle n\rangle^{1/3}}{r_{\rm vdW}^{-1}}$}+\phi\Big)\right]\langle n^{2/3}\rangle
\label{GammaA}
\end{eqnarray}
where $A\approx15.9$, $B\approx8.80$, and $\phi\approx1.61$ ---see dashed black curve in Fig.~\ref{Gamma}. 
Our numerical calculations, although largely log-periodic, are
slightly asymmetric.
The results shown in Fig.~\ref{Gamma} have roughly 40\% lower
amplitude than the experimental decay rate for $^{85}$Rb \cite{klauss2017PRL}, however, 
the oscillation phase is consistent with preliminary experimental observations \cite{klauss2017PhD}.
While our results account for losses at unitarity only, 
the experimental data was obtained after a $B$-field sweep to weak interactions \cite{klauss2017PRL}, 
thus allowing for additional atom loss. 
Nevertheless, the existence of the log-periodic oscillations in Fig.~\ref{Gamma}, with a substantial amplitude, 
violates the universality hypothesis~\cite{ho2004PRL}, in which all observables related to the unitary Bose gas should scale
continuously as powers of $n$.
Equation~(\ref{GammaA}) depends only on the system parameters $r_{\rm vdW}$ and $\eta$ for a 
particular atomic species.

We now shift our focus to the dynamical formation of weakly bound diatomic and 
Efimov states in quenched unitary Bose gases. In fact, in the recent $^{85}$Rb experiment of Ref.~\cite{klauss2017PRL} 
a population of such few-body bound states was obtained by quenching the system to unitary, evolving for a time $t_{\rm dwell}$, 
and subsequently sweeping the system back to weaker interactions ($a\approx700a_0$) within a time $t_{\rm sw}$. 
There is still, however, much to be understood on the dependence of populations on the various parameters 
($n$, $t_{\rm dwell}$, and $t_{\rm sw}$) and the possible connections to the non-equilibrium dynamics in the unitary regime.
In order to address some of these questions we 
focus initially on the case $t_{\rm sw}\ll t_{n}$, where ramping effects are minimized, and solve the time-dependent
 three-body Schr\"odinger equation 
following the same experimental protocol of Ref.~\cite{klauss2017PRL} described above ---see also 
Refs.~\cite{SupMat,baekhoj2014JPB}. 
Figure \ref{RampOut}(a) shows the three-body energy spectrum for a given density and for a range of $a$ relevant for the 
$^{85}$Rb experiment. 
For this particular density, $\langle n\rangle\approx9.9\times10^{10}$cm$^{-3}$, only the ground- 
[$\beta=0$, not shown in Fig.~\ref{RampOut}(a)] and first-excited Efimov states ($\beta=1$) have sizes smaller than the average interatomic distance. 
The black solid line in Fig.~\ref{RampOut}(a) is the energy of the (free-space) diatomic state, 
$-\hbar^2/ma^2$, while the red curves following along this threshold correspond to atom-diatom states, and those
following along the $E=0$ threshold correspond to three-atom states. 
Figure \ref{RampOut}(b) shows the population changes during the sweep of the interactions ($t_{\rm sw}\approx0.17t_n$)
from $|a|=\infty$ to $a\approx700a_0$, for a case in which the system is not held in the unitary regime ($t_{\rm dwell}=0$). 
Whenever $t_{\rm dwell}=0$, the population of few-body bound states develops entirely during the interaction sweep.
To quantify the population dynamics, we define the fraction of formed two-body states, 
$f_{\rm 2b}=(2/3)\sum_{\beta}|c_{\beta}|^2$, and Efimov states, $f_{\rm 3b}=\sum_{\beta}|c_{\beta}|^2$, after the sweep \cite{SupMat}. 
For the parameters of Fig.~\ref{RampOut}(b), we find that $f_{\rm 2b}$ and $f_{\rm 3b}$ are approximately $0.095$ and $0.004$, respectively, 
with the remaining fraction of atoms unbound. 

\begin{figure}[htbp]
\includegraphics[width=3.4in,angle=0,clip=true]{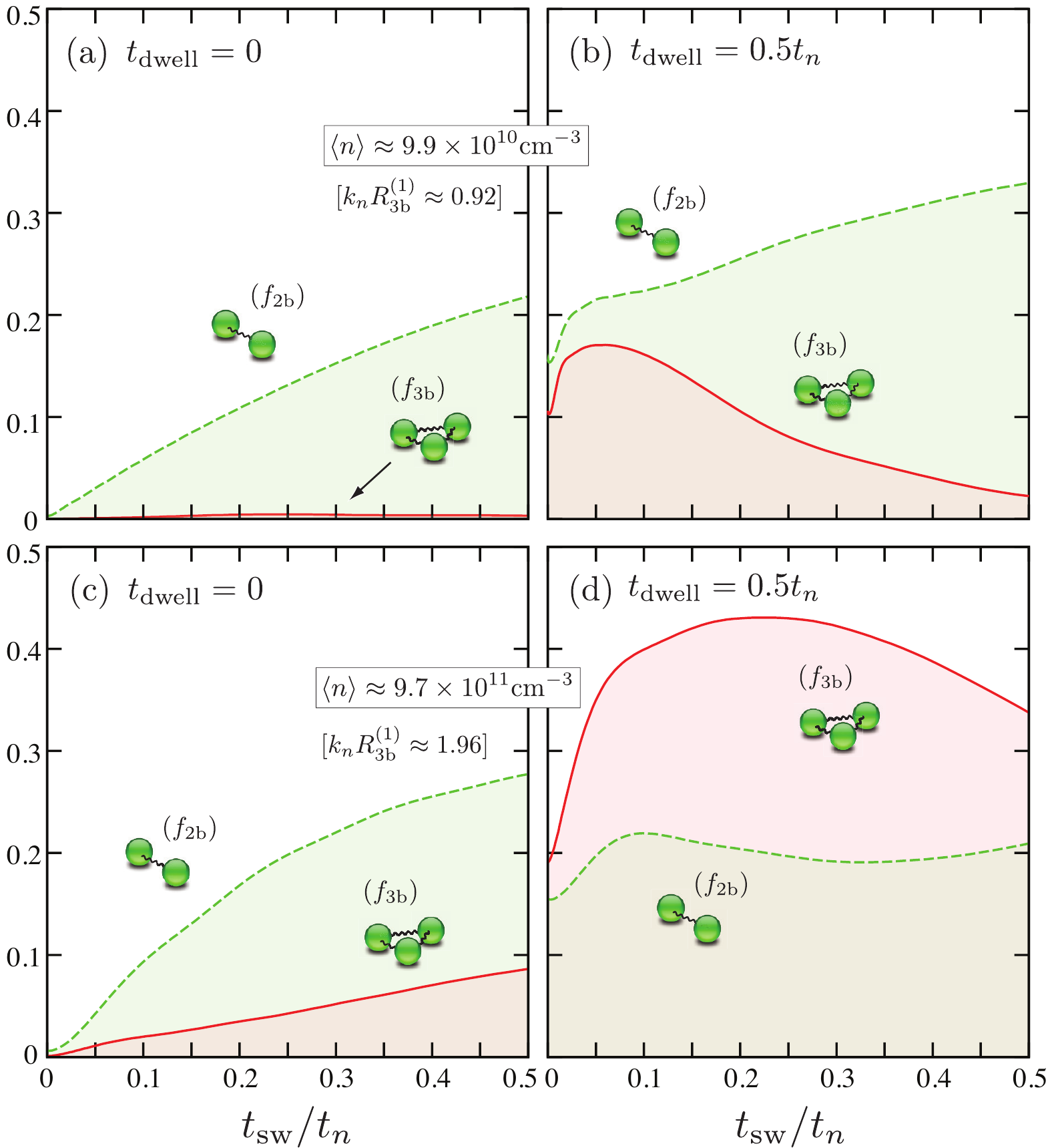}
\caption{Dependence of $f_{\rm 2b}$ and $f_{\rm 3b}$ on $t_{\rm sw}$
for (a)-(b) $\langle n\rangle\approx9.9\times10^{10}$cm$^{-3}$ and (c)-(d) $9.7\times10^{11}$cm$^{-3}$.
For $t_{\rm dwell}=0$, the dependence on $t_{\rm sw}$ is well described by the Landau-Zener results (see text)
while for $t_{\rm dwell}=0.5t_n$ [(b) and (d)] the growth of correlations lead to the enhancement of the trimer population
and a departure from the Landau-Zener results.}
\label{Ramp}
\end{figure}

To explore how the time-evolution of the system at unitarity impacts the formation of two- and 
three-body bound states, we study their dependence on $t_{\rm dwell}$ over a range of 
atomic densities [see shaded region in Fig.~\ref{EnTnCn}(b)].
Figure \ref{RampOut}(c) shows that for a relatively low density where,
although $f_{\rm 2b}$ grows fast and reaches appreciable values, $f_{\rm 3b}$ still remains 
negligible for all $t_{\rm dwell}$. A larger population of Efimov states, however, is observed for 
higher densities [see Figs. \ref{RampOut}(d) and (e), and Ref.~\cite{SupMat}.]
In general, we find that for short-times $f_{\rm 2b}\propto t_{\rm dwell}$ and $f_{\rm 3b}\propto t_{\rm dwell}^2$, 
consistent with the early-time growth of two- and three-body correlations found in Ref.~\cite{colussi2018PRL}.
Also, in Ref.~\cite{colussi2018PRL}, it was observed that at early-times the largest enhancement of three-body correlations occurred
at densities where the average interatomic distance is comparable to the size of an Efimov state. More precisely, this occured
when $k_nR_{\rm 3b}=0.74(5)$, associated with a characteristic density value
\begin{align}
\langle{n_j^*}\rangle\approx0.41\frac{(3/2)^{1/2}}{4\pi^2(1+s_0^2)^{3/2}}\frac{\kappa_*^3}{(e^{\pi/s_0})^{3j}}.\label{nc}
\end{align}
For the range of densities explored in the $^{85}$Rb experiment \cite{klauss2017PRL} the relevant characteristic density is 
$\langle{n_1^*}\rangle\approx5.2\times10^{10}$cm$^{-3}$.
In fact, Figs. \ref{RampOut} (c)-(e) display a marked change in the dynamics of $f_{\rm 3b}$ as $\langle{n_1^*}\rangle$ is 
approached from below and then exceeded. In Figs. \ref{RampOut} (c)-(e) we also display the corresponding values
for $k_nR_{\rm 3b}$.
For densities beyond $\langle{n_1^*}\rangle$ (see also Ref.~\cite{SupMat}) the population of Efimov states
is clearly enhanced and can be attributed to the growth of few-body correlations \cite{colussi2018PRL}. 
Within our model, such enhancement is also consistent with the increase of population of the first-excited Efimov 
state at unitarity [see Fig.~\ref{EnTnCn}(b)].

Correlation growth can also be further studied by investigating the dependence of the populations 
on $t_{\rm sw}$, shown in Fig.~\ref{Ramp} for two densities and for $t_{\rm dwell}=0$ and $0.5t_n$.
For $t_{\rm dwell}=0$ [Figs.~\ref{Ramp}(a) and (c)], i.e., in absence of time evolution at unitarity and the corresponding 
growth of correlations, the dependence of the populations on $t_{\rm sw}$ is well described by the Landau-Zener result: 
$f=f_m(1-e^{-t_{\rm sw}/t_m})$, where $f_{m}$ is the final population and $t_m$ the time scale related to the 
strength of the couplings between the states involved in the process \cite{stecher2007PRL,clark1979PLA}.
These findings are drastically changed as the system is allowed to evolve at unitarity, coinciding with the growth of few-body
correlations \cite{colussi2018PRL}. In that case, as shown in Figs.~\ref{Ramp}(b) and (d), the population dynamics departs from 
the Landau-Zener results. For $t_{\rm dwell}\ne0$, there is a clear enhancement of $f_{\rm 3b}$ ---even at $t_{\rm sw}=0$--- which, in 
some cases [Fig.~\ref{Ramp}(d) and Ref.~\cite{SupMat}], can lead to a population of Efimov states exceeding that of diatomic states.

In summary, solving the three-body problem in a local harmonic trap, designed to reflect density effects, 
we highlight the importance of Efimov physics in quenched unitary Bose gases. Within our model, 
the continuous scaling invariance of unitary Bose gases is violated in relevant three-body observables. 
In the three-body decay rates, this violation manifests through the appearance of log-periodic oscillations characteristic 
of the Efimov effect. In the early-time population growth of Efimov states after the system is swept away from unitarity it 
manifests through a marked change in dynamics as the density exceeds a characteristic value corresponding to a 
length scale matching that of the Efimov state size and interparticle spacing. Furthermore, our study characterizes the growth of 
correlations at unitarity through the early-time dynamics of the population of diatomic and Efimov states.  This is shown to be 
qualitatively consistent with the early-time growth of two- and three-body correlations at unitarity observed in Ref.~\cite{colussi2018PRL}. 
Moreover, we find that the departure from the Landau-Zener results for the populations 
in the non-equilibrium regime can also be associated with the increase of correlations in the system.
An experimental study of the predictions of our model is within the range of current quenched unitary Bose gas experiments.  

The authors thank E. A. Cornell, C. E. Klauss, J. L. Bohn, J. P. Corson, D. Blume and S. Kokkelmans for extensive
and fruitful discussions. J. P. D. acknowledges support from the U.S. National Science Foundation (NSF), 
Grant PHY-1607204, and from the National Aeronautics and Space Administration (NASA).
V.E.C. acknowledges support from the NSF under Grant PHY-1734006 and by Netherlands Organization for 
Scientific Research (NWO) under Grant 680-47-623.


\setcounter{equation}{0}
\setcounter{figure}{0}

\renewcommand{\theequation}{S.\arabic{equation}}
\renewcommand{\thefigure}{S.\arabic{figure}}


\section{Supplementary Material}

\subsection{A. Lifetime of three-body states}

\begin{figure}[htbp]
\includegraphics[width=3.in,angle=0,clip=true]{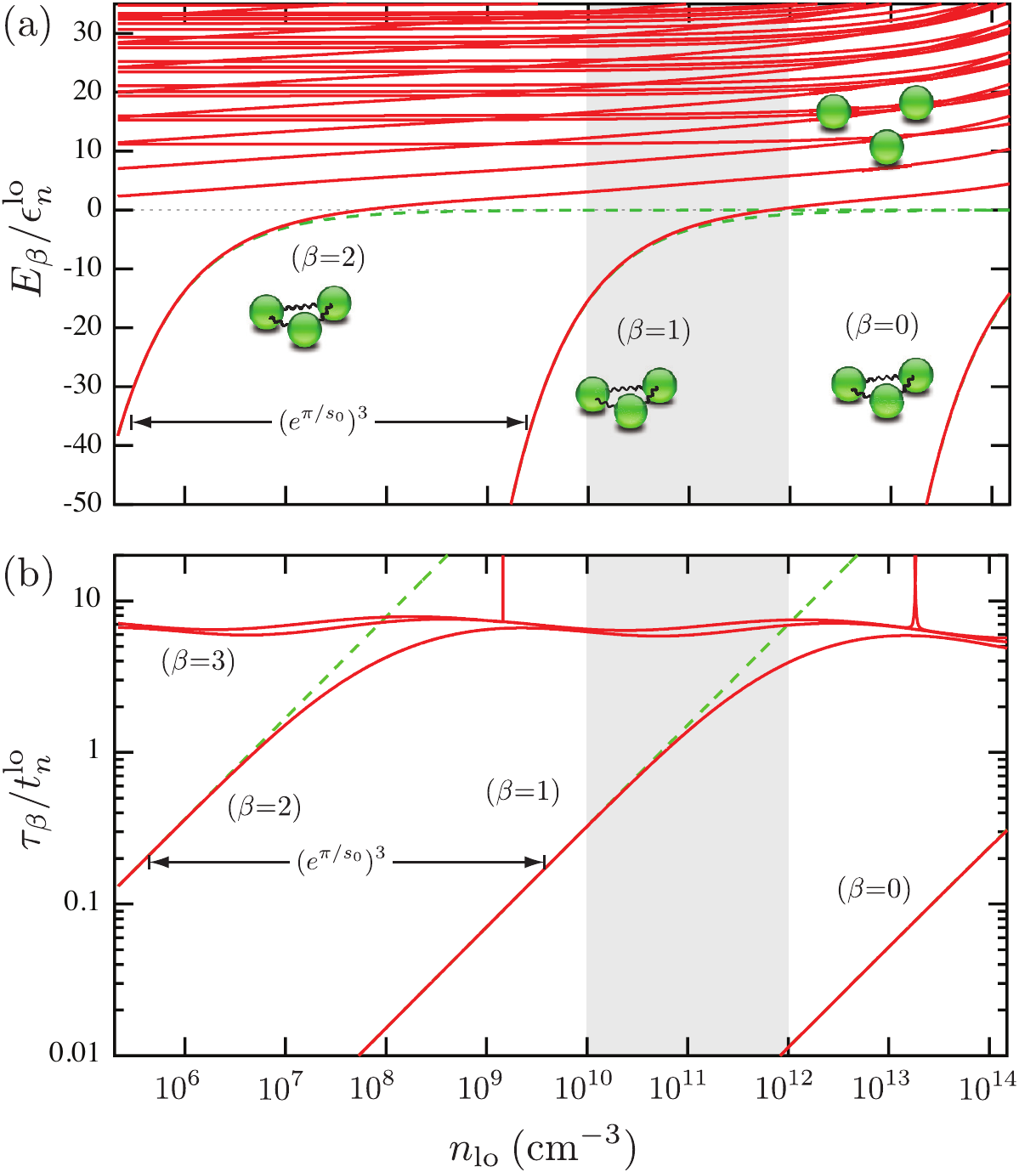}
\caption{(a) Three-body energies, $E_{\beta}$, and (b) corresponding lifetimes, $\tau_\beta$, of three-identical bosons at $|a|=\infty$ 
as a function of the local density, $n_{\rm lo}$ or, equivalently, as a function of the strength of the local harmonic trap in Eq.~(4).
The dashed green lines represent the energies and lifetimes of Efimov states in free space.}
\label{EnTn}
\end{figure}

Figure~\ref{EnTn}(a) shows the energy levels, $E_\beta$, of three-identical bosons at $|a|=\infty$ as a function 
of local density $n_{\rm lo}$, related to the local harmonic trap used in our model [see Eqs.~(3) and 
(4) of the main text]. For the present study, 
atoms interact via a Lennard-Jones potential, 
$v(r)=-C_6/r^6(1-\lambda^6/r^6)$,  where $C_6$ is the dispersion coefficient \cite{chin2010RMP}, 
and set $\lambda\approx0.9197\:r_{\rm vdW}$ to give $|a|=\infty$, such that only a single (zero-energy) $s$-wave 
dimer can exist. (For $^{85}$Rb, $r_{\rm vdW}\approx82.5a_0$ \cite{kempen2002PRL,chin2010RMP} and 
atomic mass $m\approx1.56\times10^5$a.u..) 
In that case, the width $\Gamma_\beta$ of three-body states is 
obtained by introducing a purely hyperradial absorbing potential $iV_{\rm ab}(R)/2$, 
mimicking decay to deeply bound molecular states.
The resulting width of the states can be then determined by 
\begin{align}
\Gamma_{\beta}&=\langle\psi_\beta|V_{\rm ab}(R)|\psi_\beta\rangle\nonumber\\
&=\sum_{\nu}\langle F_{\beta\nu}(R)|V_{\rm ab}(R)|F_{\beta\nu}(R)\rangle,\label{Gamma}
\end{align}
where $\psi_\beta$ is the total three-body wave function for the state 
$\beta$ and $F_{\beta\nu}$ the corresponding hyperradial amplitude for channel $\nu$ [Eqs.~(1) and (2)].

Our definition of the width in Eq.~(\ref{Gamma}) is in close analogy to other proposed ways to estimate the 
width of Efimov states \cite{braaten2006PRep,werner2006PRL} 
because it is also based on the probability to find atoms at short-distances, where inelastic transitions
to deeply bound states occur. 
In our calculations we have used $V_{\rm ab}(R)=D\exp[-R^2/(2\rho^2)]$, with the parameters $D$ and $\rho$ adjusted 
to reproduce the expected width via the relation
\begin{align}
\Gamma_0=\frac{4\eta}{s_0}|E_{0}|, \label{GammaRes}
\end{align}
where $E_{0}$ is the energy of the Efimov ground state in free space at $|a|=\infty$. 
In our present study we assume $\eta=0.06$. This is consistent with the observed value for $^{85}$Rb 
\cite{wild2012PRL}, and vary $\rho$ and $D$ until Eq.~(\ref{GammaRes}) is satisfied. 
This is achieved for 
the choice of parameters: $\rho=2r_{\rm vdW}$ and  $D=1.8\times10^{-1}\hbar^2/mr_{\rm vdW}^2$. 
The resulting lifetimes $\tau_\beta=\hbar/\Gamma_\beta$ of the three-identical bosons
states from Fig.~\ref{EnTn}(a) are shown in Fig.~\ref{EnTn}(b). Away from 
$|a|=\infty$, the same parameters $\rho$ and $D$ lead to a lifetime of $109\mu$s for the first-excited Efimov state, 
consistent with the observations of Ref.~\cite{klauss2017PRL}, indicating the validity of our decay model [Eq.~(\ref{Gamma})]
for other values of $a$. 

\subsection{B. Field sweep of interactions}

\begin{figure}[htbp]
\includegraphics[width=3.in,angle=0,clip=true]{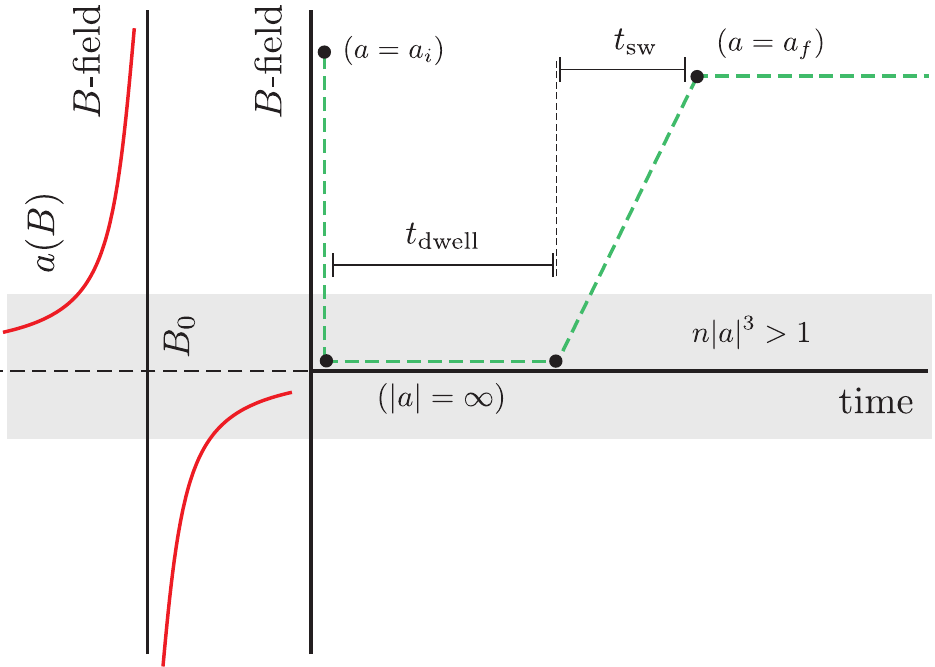}
\caption{Magnetic field ramp scheme implemented in the $^{85}$Rb experiment. 
Initially ($t=0$), the system is rapidly quenched by quickly changing the magnetic field from $a\approx150a_0(=a_i)$ 
to $|a|=\infty$. The the system is 
held at unitarity for a variable time $t_{\rm dwell}$. Finally, the system is slowly swept away from unitarity to weaker 
interactions ($a_f\approx700a_0$) over a time $t$.}
\label{RampScheme}
\end{figure}

In the $^{85}$Rb experiment with unitary Bose gases \cite{klauss2017PRL}, a sequence of magnetic field ramps
has been implemented in order to study the dynamics of the system. Reference \cite{klauss2017PRL} uses the
155G Feshbach resonance with parameters: $B_0\approx155.041$G, $a_{bg}\approx-443a_0$ and $\Delta\approx10.71$G
\cite{claussen2003PRA}, where $B_0$ is the resonance position, $a_{bg}$ the background scattering length and $\Delta$ the resonance
width. The ramping scheme is shown in Fig.~\ref{RampScheme}. At $t=0$ the system is initially in the weakly interacting ground state, 
$B$ is then quickly ramped from a value where $a\approx150a_0(=a_i)$ ($B\approx B_0+8G$)
to a value in which $|a|=\infty$ ($B=B_0$), thus quenching the interactions. 
The shaded region in Fig.~\ref{RampScheme} represents the range of $B$ where $n|a|^3>1$.
The system is then held at unitarity for a variable time $t_{\rm dwell}$ and then slowly
swept away to weaker interactions ($a_f\approx700a_0$, or $B\approx159$G) over a time $t_{\rm sw}$. 
For the Lennard-Jones potential we use (see above) the values of $\lambda$ producing
$a_i\approx0.7725\:r_{\rm vdW}$ and $a_f\approx0.8922\:r_{\rm vdW}$.
Once the system is swept away from unitarity, the population of weakly bound diatomic and Efimov states is determined \cite{klauss2017PRL}.

There are three distinct regions in Fig.~\ref{RampScheme} where the three-body physics is studied within our model. 
For $t<0$, the three-body system is in the lowest trapped eigenstate of the Hamiltonian 
for weak interactions ($a=a_i$) [lowest positive energy state on the far right of Fig.~3(a)].
The total wave function for $t<0$ is given by
\begin{eqnarray}
\Psi_{i}(R,\Omega,t)=\psi_{i}(R,\Omega)e^{-iE_{i}t/\hbar}.
\end{eqnarray}
At $t=0$, the interaction is quenched and the total wave function
is given by a superposition of trapped three-body states at unitarity [see states on the left of Fig.~3(a)],
\begin{align}
\Psi_u(R,\Omega,t)=\sum_\beta c_{\beta}^{(u)}\psi_\beta^{(u)}(R,\Omega)e^{-i(E_{\beta}^{(u)}-i\Gamma_{\beta}^{(u)}/2)t/\hbar},\label{PsiU}
\end{align}
with coefficients $c_\beta$ 
\begin{align}
c_\beta^{(u)}&=\big\langle\psi^{(u)}_{\beta}\big|\Psi_i(t=0)\big\rangle\nonumber\\
&=\sum_{\nu\nu'}\int [F_{\nu\beta}^{(u)}(R)]^*F_{\nu'\beta_i}^{(i)}(R)  dR\nonumber\\
&~~~~~~\times\int  [\Phi_{\nu}^{(u)}(R;\Omega)]^*\Phi_{\nu'}^{(i)}(R;\Omega) d\Omega.\label{CbetaU}
\end{align}
This determines the time evolution of the system until $t=t_{\rm dwell}$.
For $t_{\rm dwell}\le t\le t_{\rm dwell}+t_{\rm sw}$, interactions are slowly swept from $|a|=\infty$ to $a_f$ 
[from left to right of Fig. 3(a)], and the time evolution of the system
is dictated by the time-dependent Schr\"odinger equation:
\begin{eqnarray}
H(R,\Omega,t)\Psi_{\rm sw}(R,\Omega,t)=i\hbar\frac{\partial}{\partial t}\Psi_{\rm sw}(R,\Omega,t). \label{SchroEq}
\end{eqnarray}
For $t\ge t_{\rm dwell}+t_{\rm sw}$, the time evolution is given by
\begin{align}
\Psi_f&(R,\Omega,t)=\nonumber\\
&\sum_\beta c_{\beta}^{(f)}\psi_\beta^{(f)}(R,\Omega) e^{-i(E_{\beta}^{(f)}-i\Gamma_{\beta}^{(f)}/2)(t-t_{\rm dwell}-t_{\rm sw})/\hbar}.\label{PsiF}
\end{align}
We then determine the final state populations resulting from the ramping scheme in Fig.~\ref{RampScheme} from
\begin{align}
|c_\beta^{(f)}|^2&=|\big\langle\psi^{(f)}_{\beta}\big|\Psi_{\rm sw}(t=t_{\rm dwell}+t_{\rm sw})\big\rangle|^2.
\end{align}

\subsection{C. SVD approximation in time}

Numerically solving the time-dependent Schr\"odinger equation
(\ref{SchroEq}) is crucial to studying the ramping scheme in Fig.~\ref{RampScheme}.
However, from the variation of the energy spectrum in Fig. 3 in time as $a$ changes, it is clear that an 
adiabatic approach in time will encounter numerical instabilities due to sharp avoid-crossings. In order to circumvent 
such issues, we extend the SVD approach of Ref.~\cite{wang2011PRA} to the time domain, in a similar fashion to 
Ref.~\cite{baekhoj2014JPB}.

In our SVD formulation in the time domain, the total wave function [solution of Eq.~(\ref{SchroEq})]
is expanded in terms of the DVR basis \cite{lill1982CPL,rescino2000PRA}, $\pi_{j}(t)$, in the following way:
\begin{eqnarray}
\Psi(R,\Omega,t)=\sum_{j\beta}C_{j\beta} \pi_j(t)\psi^{(t_j)}_{\beta}(R,\Omega),\label{PsiSVD}
\end{eqnarray}
where $\psi^{(t)}_{\beta}$ is the eigenstate of $H$ at time $t$, with energy given by $E_{\beta}(t)$. 
We determine the $C_{j\beta}$ coefficients for a given initial condition. 
Various properties of the DRV basis derived below are needed to derive the form of the Hamiltonian matrix elements \cite{wang2011PRA}.

The DVR basis functions $\pi_j(t)$ are defined by the Gauss-Lobatto quadrature points $x_j$ and weights $w_j$ 
\cite{manolopoulos1998cpl}. The Gauss-Lobatto quadrature approximates integrals of a function $g\left( t \right)$ as
\begin{equation}\label{LobattoQuadrature}
\int_{t_1}^{t_N} g(t)dt  \cong \sum_{j = 1}^{N} g(t_j) \widetilde{w}_j,
\end{equation}
where $N$ is the number of quadrature points and
\begin{equation}
\widetilde{w}_j=\frac{t_N-t_1}{2}w_j,~~~t_j=\frac{t_N-t_1}{2}x_j +\frac{t_N+t_1}{2}.
\end{equation}
Equation (\ref{LobattoQuadrature}) is exact for polynomials of degree less than or equal to $2N-1$. 
The DVR basis functions and their derivatives are constructed based on values for $t_j$ and 
$\widetilde{w}_j$ as following
\begin{eqnarray}\label{DVR}
\pi_j(t)&=&\sqrt{\frac{1}{\widetilde{w}_j}}\prod\limits_{k \ne j}^N {\frac{{t - t_k }}{{t_j  - t_k }}},\\
\frac{d}{dt}\pi_j(t)&=&\sqrt{\frac{1}{\widetilde{w}_j}}\sum_{l\ne j}^{N}\bigg(\frac{1}{t-t_l}
\prod\limits_{k \ne j,l}^N {\frac{{t - t_k }}{{t_j  - t_k }}}\bigg).
\end{eqnarray}
This leads to important properties for calculating matrix elements at the quadrature points,
\begin{eqnarray}\label{DVRProperty}
\pi _j(t_i)&=&\sqrt{\frac{1}{\widetilde{w}_j}}\delta _{ij}~~~(\forall~i,j),\\
\pi _j'(t_i)&=&\sqrt{\frac{1}{\widetilde{w}_j}}\frac{1}{t_j-t_i}\prod\limits_{k \ne i,j}^N {\frac{{t_i - t_k }}{{t_j  - t_k }}}~~~(i\ne j),\\
\pi _j'(t_i)&=&\sqrt{\frac{1}{\widetilde{w}_j}}\frac{\delta_{jN}-\delta_{j1}}{2\widetilde{w}_j}~~~(i=j).
\end{eqnarray}
Property (\ref{DVRProperty}), for instance, can be used to calculate integrals involving the DVR basis and an arbitrary function 
$F(t)$ in a trivial way,
\begin{eqnarray}
\int_{t_a}^{t_b}\pi_j(t)F(t)\pi_k(t)dt&\cong&\sum_{l=1}^{N}\pi_j(t_l)F(t_l)\pi_k(t_l)\widetilde{w}_l\nonumber\\
&=&F(t_j)\delta_{jk},
\end{eqnarray}
which is usually called the DVR approximation.

Now, substituting Eq.~(\ref{PsiSVD}) in Eq.~(\ref{SchroEq}), and projecting out in the basis $\pi_i(t)\psi_{\alpha}^{(t_i)}$,
yields a linear system of equations for the coefficients $C_{j\beta}$,
\begin{eqnarray}
\sum_{j\beta}\left(\widetilde{H}_{i\alpha,j\beta}-\widetilde{L}_{i\alpha,j\beta}\right)C_{j\beta}=0.\label{HLEq}
\end{eqnarray}
Equivalently, this system of equations can be written in the matrix form $\widetilde{H}\vec{C}=\widetilde{L}\vec{C}$, with matrix elements
given by
\begin{eqnarray}
\widetilde{H}_{i\alpha,j\beta}&=&E_\beta(t_i)\delta_{\alpha\beta}\delta_{ij}\nonumber\\
&-&\frac{i\hbar}{2}\int_{t_a}^{t_b}\left[\pi_i(t)\frac{d}{dt}\pi_j(t)-\frac{d}{dt}\pi_i(t)\pi_j(t)\right]dt~O^{i\alpha}_{j\beta}\nonumber\\
&=&E_\beta(t_i)\delta_{\alpha\beta}\delta_{ij}\nonumber\\
&-&\frac{i\hbar}{2}\left[\sqrt{\widetilde{w}_i}\pi'_j(t_i)-\sqrt{\widetilde{w}_j}\pi'_i(t_j)\right]O^{i\alpha}_{j\beta},\label{Hmat}
\end{eqnarray}
and
\begin{eqnarray}
\widetilde{L}_{i\alpha,j\beta}&=&\frac{i\hbar}{2}\left[\pi_i(t_b)\pi_j(t_b)-\pi_i(t_a)\pi_j(t_a)\right]O^{i\alpha}_{j\beta}\nonumber\\
&=&\frac{i\hbar}{2}\left(\frac{\delta_{iN}\delta_{jN}}{\widetilde{w}_N}-\frac{\delta_{i1}\delta_{j1}}{\widetilde{w}_1}\right)\delta_{\alpha\beta},\label{Lmat}
\end{eqnarray}
where the overlapping matrix is calculated from
\begin{eqnarray}
O^{i\alpha}_{j\beta}=\int\psi^{(t_i)*}_{\alpha}(R,\Omega)\psi^{(t_j)}_{\beta}(R,\Omega)d\Omega dR.\label{Omat}
\end{eqnarray}

This problem can be solved by writing Eq.~(\ref{HLEq}) as
$\vec{C}=(\widetilde{H}^{-1}L)\vec{C}$. Due to the form of the matrix $L$ [Eq.~(\ref{Lmat})], however,
the coefficients $\vec{C}$ can be written as
\begin{eqnarray}
C_{i\alpha}&=&\sum_{j\beta}\left(\widetilde{H}^{-1}L\right)_{i\alpha,j\beta}C_{j,\beta}\nonumber\\
&=&\sum_{\beta}(R_{iN})_{\alpha\beta}C_{N\beta}-\sum_{\beta}(R_{i1})_{\alpha\beta}C_{1\beta},\label{Ccoefs}
\end{eqnarray}
where $R_{ij}$ is the matrix defined as
\begin{eqnarray}
(R_{ij})_{\alpha\beta}=\frac{i\hbar}{2\widetilde{w}_j}(\widetilde{H}^{-1})_{i\alpha,j\beta}.
\end{eqnarray}
Now, writing Eq.~(\ref{Ccoefs}) in the matrix notation,
\begin{eqnarray}
\vec{C}_{i}&=&R_{iN}\vec{C}_{N}-R_{i1}\vec{C}_{1},\label{CmatX}
\end{eqnarray}
where $\vec{C}_i$ is a vector of components $C_{i\alpha}$, with $\vec{C}_{1}$ determined by the initial condition (see below)
and $\vec{C}_{N}$ by 
\begin{eqnarray}
\vec{C}_{N}&=&(R_{1N})^{-1}(\mathbb{1}+R_{11})\vec{C}_{1},\label{CN}
\end{eqnarray}
obtained by making $i=1$ in Eq.~(\ref{CmatX}).
Putting together these results we finally obtain
\begin{eqnarray}
\vec{C}_{i}&=&\left[R_{iN}(R_{1N})^{-1}(\mathbb{1}+R_{11})-R_{i1}\right]\vec{C}_{1},\label{Cmat}
\end{eqnarray}
determining the coefficients $C_{i\beta}$ in Eq.~(\ref{PsiSVD}) for all grid points 
in the interval $t \in \left[t_1,t_N\right]$. An alternative solution can
be obtained by setting $i=N$ in Eq.~(\ref{CmatX}). In that case, we obtain
\begin{eqnarray}
\vec{C}_{i}&=&-\left[R_{iN}(\mathbb{1}-R_{NN})^{-1}R_{N1}+R_{i1}\right]\vec{C}_{1}.\label{Cmat2}
\end{eqnarray}
We have observed that the numerical calculations are more stable by choosing the propagation
scheme defined by Eq.~(\ref{Cmat2}).

\begin{figure*}[htbp]
\includegraphics[width=5.4in,angle=0,clip=true]{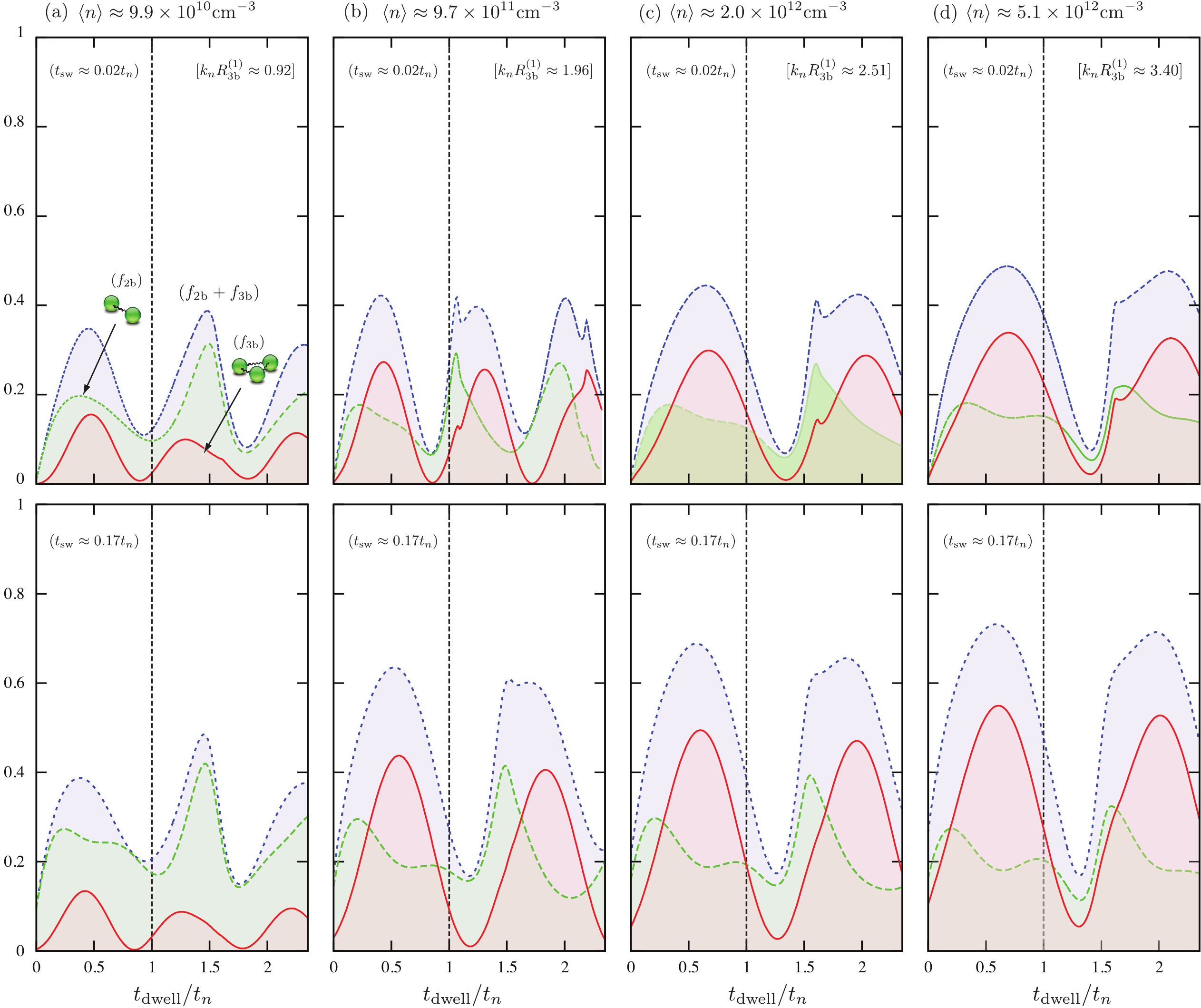}
\caption{Dependence of $f_{\rm 2b}$, $f_{\rm 3b}$, and $f_{\rm 2b}+f_{\rm 3b}$, on $t_{\rm dwell}$ for various
densities, satisfying the condition $k_nR_{\rm 3b}>0.74(5)$ \cite{colussi2018PRL}, and $t_{\rm sw}=0.02t_n$ and $0.17t_n$ 
(top and bottom panels, respectively). 
In all cases, the $t_{\rm dwell}$ dependence is shown for times beyond the range of validity 
of our approach ($t<t_n$) to illustrate the time periodicity implied by our local harmonic trap.}
\label{RampOutSup}
\end{figure*}

Applying this method to the current problem, we find that dividing the time domain into sectors is 
advantageous. 
Within the first sector, the ${C}_{1\beta}$ coefficients 
correspond to the initial condition at $t=t_{\rm dwell}$ determined by simply equating Eq.~(\ref{PsiU}) and 
Eq.~(\ref{PsiSVD}) and making use the property of the DVR basis
in Eq.~(\ref{DVRProperty}). This leads to
\begin{eqnarray}
C_{1\beta}=\sqrt{\widetilde{w}_1} c_{\beta}^{(u)} e^{-i(E_{\beta}^{(u)}-i\Gamma_{\beta}^{(u)})t_{\rm dwell}/\hbar}.
\end{eqnarray}
The coefficients corresponding to the last point within the first sector, $C_{N\beta}$, serve as the initial condition 
for the propagation within the next sector. This procedure continues until the last sector, with the last point corresponding 
to $t_N=t_{\rm dwell}+t_{\rm sw}$. From this point, we determine the coefficients of the final total wavefunction in Eq.~(\ref{PsiF}) as
\begin{align}
c_{\beta}^{(f)}=C_{N\beta}/\sqrt{\widetilde{w}_N}.
\end{align}
As a result, the final population of a state $\beta$ is simply determined by calculating $|c_{\beta}|^2$.
In order to associate these populations to the actual populations of diatomic and triatomic 
states, the nature of the particular $\beta$ states must be accounted. For instance, $\beta$ states
whose energies follow the black solid line in Fig.~3(a) ($E_{\beta}>-\hbar^2/ma^2$) represent
an atom-diatom state. In that case, the corresponding population of diatomic molecules is given by
$f_{\rm 2b}=(2/3)\sum_{\beta}|c_{\beta}|^2$. For triatomic states, with energies $E_\beta<-\hbar^2/ma^2$, 
the population is determined by $f_{\rm 3b}=\sum_{\beta}|c_{\beta}|^2$. 

\subsection{D. Populations of diatomic and Efimov states}

\begin{figure*}[htbp]
\includegraphics[width=5.4in,angle=0,clip=true]{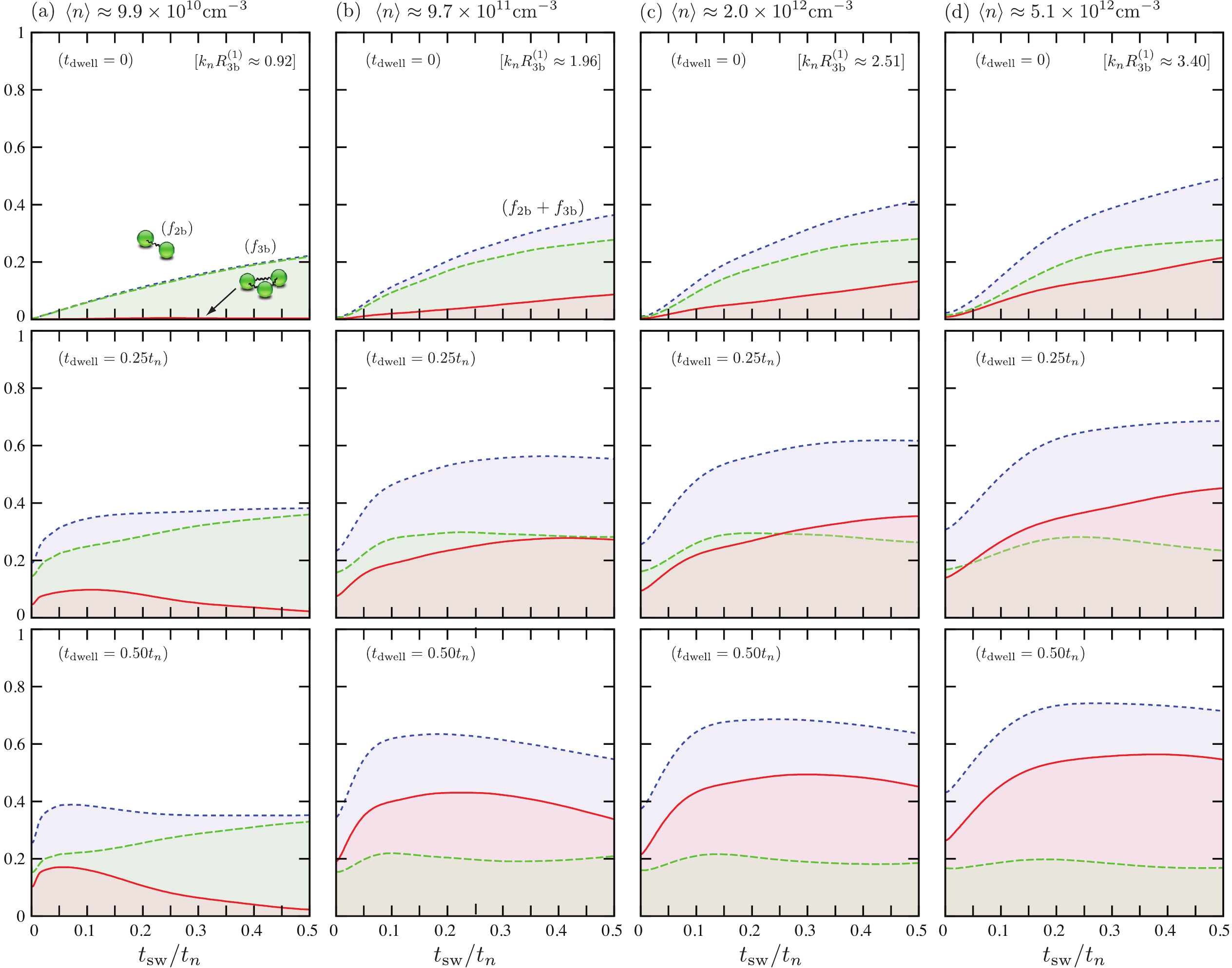}
\caption{Dependence of $f_{\rm 2b}$, $f_{\rm 3b}$, and $f_{\rm 2b}+f_{\rm 3b}$, on $t_{\rm sw}$ 
for various densities and $t_{\rm dwell}=0$, $0.25t_n$, and $0.50t_n$ (top, middle, and bottom panels, respectively).}
\label{RampOutTswSup}
\end{figure*}

Figures \ref{RampOutSup} and \ref{RampOutTswSup} display various calculations for the population of weakly bound 
diatomic and Efimov states that support and complement the conclusions of the main text. In Fig. \ref{RampOutSup}
we show the dependence of $f_{\rm 2b}$, $f_{\rm 3b}$, and $f_{\rm 2b}+f_{\rm 3b}$, on $t_{\rm dwell}$ for two values of $t_{\rm sw}$ and
densities $k_nR^{(1)}_{\rm 3b}>0.74(5)$, exceeding the density where maximal early-time enhancement of three-body correlations was observed 
in Ref.~ \cite{colussi2018PRL}.  In all cases shown in Fig.~\ref{RampOutSup} we have displayed the $t_{\rm dwell}$ 
dependence for times beyond the range of validity of our approach ($t<t_\mathrm{n}$) in order to illustrate
the time periodicity due to the local harmonic trap in our model. As expected for short sweep times, our results for $t_{\rm sw}=0.02t_n$ 
[top panels of Figs.~\ref{RampOutSup}(a)-(d)] clearly show that populations are negligible for $t_{\rm dwell}\ll t_n$. However, for a finite
value of $t_{\rm dwell}$ the populations of both diatomic and triatomic states become substantial after evolving at unitary while correlations
develop \cite{colussi2018PRL}. For longer sweep times, $t_{\rm sw}=0.17t_n$ 
[bottom panels of Figs.~\ref{RampOutSup}(a)-(d)], a more substantial population is observed for $t_{\rm dwell}\ll t_n$. This indicates
that $t_{\rm sw}=0.17t_n$ is slow enough to produce populations even if the system is does not evolve at 
unitarity.

Figure \ref{RampOutTswSup} shows the dependence of $f_{\rm 2b}$ and $f_{\rm 3b}$, as well as $f_{\rm 2b}+f_{\rm 3b}$, on $t_{\rm sw}$ 
for various densities and three values of $t_{\rm dwell}$. These results complement our discussion on the breakdown of the Landau-Zener 
model in the main text. For a two-level system, the Landau-Zener model predicts that after sweeping the interactions across an avoided-crossing the final state
population is determined by $f=f_m(1-e^{-t_{\rm sw}/t_m})$, where $f_{m}$ is the final population and $t_m$ the time scale related to the 
strength of the couplings between the states involved in the process \cite{stecher2007PRL,clark1979PLA}. The results for $t_{\rm dwell}=0$, 
displayed in the top panels of Figs.~\ref{RampOutTswSup}(a)-(d), agree well with the Landau-Zener formula ---this agreement tends to 
deteriorate as the density increases. For $t_{\rm dwell}\ne0$ [middle and bottom panels of Figs.~\ref{RampOutTswSup}(a)-(d)], however, 
our results depart from the Landau-Zener formula. This departure coincides with the growth of few-body
correlations \cite{colussi2018PRL}, thus providing an alternative way to characterize such phenomena.

\begin{thebibliography}{10}
\bibitem{SupMat}
``Url will be inserted by publisher,''
\end{thebibliography}

\end{document}